\documentclass[aps,pra,twocolumn,superscriptaddress,10pt,tightenlines,nofootinbib]{revtex4}
\usepackage{amsmath,amsthm,graphicx,amssymb,verbatim,listings,mathrsfs}
\usepackage{verbatimbox}
\usepackage[usenames, dvipsnames]{color}
\usepackage[ pdftex, plainpages = false, pdfpagelabels,
                 pdfpagelayout = useoutlines,
                 bookmarks,
                 bookmarksopen = true,
                 bookmarksnumbered = true,
                 breaklinks = true,
                 linktocpage=all,
                 pagebackref=false,
                 colorlinks = true,
                 linkcolor = BrickRed,
                 urlcolor  = blue,
                 citecolor = BrickRed,
                 anchorcolor = green,
                 hyperindex = true,
                 hyperfigures
                 ]{hyperref}
\usepackage{xifthen} 

\newcommand{\ket}[1]{{\ensuremath{\left| #1 \right\rangle}}}

\newcommand{\braket}[2]{{\ensuremath{\left\langle #1 \middle| #2
      \right\rangle}}}
\newcommand{\ketbra}[2]{{\ensuremath{\left| #1 \middle\rangle \middle\langle #2
      \right|}}}

\newcommand{\arxiv}[2][]{\ifthenelse{\isempty{#1}}{\href{http://arxiv.org/abs/#2}{{\tt arXiv:\allowbreak{}#2}}} {\href{http://arxiv.org/abs/#2}{{\tt arXiv:\allowbreak{}#2 [#1]}}}}

\newcommand{\booktitle}{\textsl}
\newcommand{\hrefdoi}[2]{\href{https://dx.doi.org/#1}{#2}}

\begin{document}

\title{Sporadic SICs and the Normed Division Algebras}

\author{Blake C.\ Stacey}
\affiliation{Department of Physics,
  University of Massachusetts Boston, 100 Morrissey Blvd.,
  Boston, MA 02125, United States}

\date{\today}

\begin{abstract}
Recently, Zhu classified all the SIC-POVMs whose symmetry groups act
doubly transitively.  Lattices of integers in the complex numbers, the
quaternions and the octonions yield the key parts of these symmetry
groups.
\end{abstract}

\maketitle

The problem of the SIC-POVMs~\cite{Zauner99, Renes04, Scott10} has a
classic feel: It is easy to state, it continues to prove fiendishly
hard to solve, and it makes unforeseen connections between disparate
subjects.  A symmetric, informationally complete, positive-operator
valued measure---a SIC-POVM, or just a SIC---is a set of $d^2$ vectors
$\ket{\psi_j}$ in a $d$-dimensional Hilbert space such that
\begin{equation}
\left|\braket{\psi_j}{\psi_k}\right|^2
 = \frac{d\delta_{jk} + 1}{d+1}.
\end{equation}
In practice, it is often convenient to consider the equivalent set of
rank-1 projectors, $\Pi_j = \ketbra{\psi_j}{\psi_j}$.  Either way,
such a set defines a quantum measurement operation with interesting
properties~\cite{RMP, Fuchs2014, Voldemort, ConicalDesigns, RCF-SIC,
  Zhu2016}.  The question is whether SICs exist for all values of~$d$.
The growing list of both exact and high-precision numerical
solutions~\cite{ConicalDesigns, RCF-SIC} is encouraging, but as yet,
we have no proof one way or the other.

A SIC is \emph{group covariant} if it can be constructed by starting
with a single vector (the \emph{fiducial}) and acting upon that vector
with the elements of some group.  All known SICs are group covariant,
although since group covariance simplifies the search process, this
could be a matter of the light being under the lamppost.  Furthermore,
in all cases but one, that group is a \emph{Weyl--Heisenberg group.}
Working in dimension $d$, let $\omega_d = e^{2\pi i / d}$, and define
the shift and phase operators
\begin{equation}
X\ket{j} = \ket{j+1},\ Z\ket{j} = \omega_d^j \ket{j},
\end{equation}
where the shift is modulo $d$.  Products of powers of $X$ and $Z$,
together with dimension-dependent phase factors that we can neglect
for the present purposes, define the Weyl--Heisenberg group.

In $d = 2$, we can draw a SIC in the Bloch representation.  Any qubit
SIC forms a tetrahedron inscribed in the Bloch sphere~\cite{Renes04}.
Two SICs in higher dimensions will be important for our purposes.
First is the \emph{Hesse SIC} in $d = 3$, constructed by applying the
Weyl--Heisenberg group to the fiducial
\begin{equation}
\ket{\psi_0^{\rm (Hesse)}} 
 = \frac{1}{\sqrt{2}} (0, 1, -1)^{\rm T}.
\label{eq:hesse-fiducial}
\end{equation}
Second is the \emph{Hoggar SIC} in $d = 8$.  We have multiple choices
of fiducial in this case, but they all yield structures that are
equivalent up to unitary or antiunitary transformations, so for
brevity we speak of ``the'' Hoggar SIC~\cite{zhu-thesis}.  One viable
fiducial~\cite{Szymusiak2015} is
\begin{equation}
\ket{\psi_0^{\rm (Hoggar)}} \propto (-1+2i, 1, 1, 1,
                      1, 1, 1, 1)^{\rm T}.
\label{eq:hoggar-fiducial}
\end{equation}
The Hoggar SIC is the \emph{only} known case where the group that
constructs the SIC from the fiducial is not the Weyl--Heisenberg group
for $d$ dimensions itself~\cite{zhu-thesis}.  Instead, we use the
tensor product of three copies of the qubit Weyl--Heisenberg group.

The SICs in dimensions 2 and 3, as well as the Hoggar SIC in dimension
8, stand apart in some respects from the other known
solutions~\cite{RCF-SIC, stacey-qutrit}.  They lie outside the
algebraic number theory framework of Appleby \emph{et
  al.}~\cite{RCF-SIC}.  Either their dimensions are too small, or (in
the case of the Hoggar SIC) they have the wrong symmetry group.  We
can think of them as the \emph{sporadic SICs.}  This list encompasses
all of the SICs whose symmetry groups act doubly transitively on their
projectors: the qubit SICs, the Hesse SIC in dimension 3, and the
Hoggar SIC~\cite{zhu2015}.  (The other SICs in $d = 3$ do not have
doubly-transitive symmetry groups, but should still be counted as
sporadic~\cite{RCF-SIC}.)  I will now relate the doubly-transitive
SICs with the complex numbers $\mathbb{C}$, the quaternions
$\mathbb{H}$ and the octonions $\mathbb{O}$.

The \emph{Eisenstein integers}~\cite{baez-review} are complex numbers
of the form
\begin{equation}
z = a + b\omega, \hbox{ where } \omega = e^{2\pi i / 3}
\hbox{ and } a,b \in \mathbb{Z}.
\end{equation}
In the complex plane, they form a hexagonal lattice, designated $A_2$.
A \emph{unit} among integers is an integer whose multiplicative
inverse is also an integer.  Within the familiar set $\mathbb{Z}$, we
have only two: namely, $+1$ and $-1$.  However, in the Eisenstein
integers, there exist more choices.  The group of units in the ring of
Eisenstein integers is
\begin{equation}
\{ \pm 1, \pm \omega, \pm \omega^2 \}.
\end{equation}
Note the presence of the geometrical operations that take an
equilateral triangle to itself: We see the identity, rotation by $1/3$
of a circle, and rotation by $2/3$ of a circle.  So, modulo some
signs, we have the symmetry operations that rotate a regular tetrahedron
around the axis of one vertex, holding that vertex fixed.

A qubit SIC is a tetrahedron inscribed in the Bloch sphere, and
unitary operations on qubit state space are rotations of that sphere.
Therefore, the unitaries which hold one vertex of a SIC fixed and
permute the other three form a group that is isomorphic to
$\mathbb{Z}_3$.

In other words, the \emph{stabilizer group} for each projector in a
tetrahedral SIC is \emph{the unit group of the Eisenstein integers,}
quotiented by a small simple group.

Moving to the next normed division algebra, what about the quaternions
$\mathbb{H}$?  We can define a set of integers for~$\mathbb{H}$, the
so-called \emph{Hurwitz integers}~\cite{baez-review}.  These are the
quaternions whose coefficients are either all integers in~$\mathbb{Z}$
or all half-integers.  The units of the Hurwitz integers
form the ``binary tetrahedral group''~\cite{baez-review, wilson2009}.
Thought of geometrically, the 24 units of the Hurwitz integers are the
vertices of a polytope, the \emph{24-cell,} and they are the root
vectors of the $D_4$ lattice.  The binary tetrahedral group is known to be
isomorphic to the matrix group $SL(2,3)$.

And $SL(2,3)$ is isomorphic to the stabilizer group for each of the
projectors in the Hesse SIC.
Therefore, we can say that the stabilizer of any element in the Hesse
SIC is given by the unit group of the Hurwitz integers.

The octonions $\mathbb{O}$ also have integers among
them~\cite{wilson2009, integral-octonions}.  The construction relevant
for our purposes is known as the \emph{Cayley integers.}  These form a
lattice, specifically, the $E_8$ lattice scaled by a factor
$1/\sqrt{2}$.  Exactly 240 elements in the Cayley integers have unit
norm; they correspond to the root vectors of the $E_8$ lattice.  Do
they form a group of units, as we saw in~$\mathbb{C}$ and
in~$\mathbb{H}$?  Not exactly, because the octonions are not
associative.

However, we can still avail ourselves of a group structure.  To do so,
we define an \emph{automorphism} of the octonions as an invertible
linear map from $\mathbb{O}$ to~$\mathbb{O}$ that preserves the
multiplication structure.  The automorphism group of the integral
octonions has order 12096~\cite{wilson2009}, and is sometimes written
$G_2(\mathbb{Z})$. The structure of~$G_2(\mathbb{Z})$ has been worked
out, and is given by
\begin{equation}
G_2(\mathbb{Z}) \cong PSU(3,3) \rtimes \mathbb{Z}_2.
\end{equation}
Zhu~\cite{zhu2015} identified $PSU(3,3)$ as isomorphic to the stabilizer
of each projector in the Hoggar SIC.

In summary, then: For each of the normed division algebras
$\mathbb{C}$, $\mathbb{H}$ and $\mathbb{O}$, we can build a set of
integers whose symmetries are, up to factors of~$\mathbb{Z}_2$, the
stabilizer groups for projectors in the tetrahedral SICs, the Hesse
SIC and the Hoggar SIC.  (The factors of~$\mathbb{Z}_2$ in dimensions
2 and 8 can be accounted for by considering the pair of a fiducial
with its counterpart in a \emph{twinned} SIC, related to the original
SIC by an antiunitary operation~\cite{Szymusiak2015}.  Note that
Eq.~(\ref{eq:hesse-fiducial}) is invariant under complex conjugation,
but Eq.~(\ref{eq:hoggar-fiducial}) is not.)  Recall that we noticed
the appearance of the $A_2$, $D_4$ and $E_8$ lattices.  So, we can say
that the doubly transitive SICs fall into an \emph{ADE
  classification}~\cite{ncatlab-ade}.  The occurrence of these
lattices, moreover, connects the SIC question to the problem of sphere
packing~\cite{viazovska2016}, another topic in which the solutions in
one dimension can be vexingly unrelated to solutions for others.


\begin{thebibliography}{999}
\bibitem{Zauner99} G. Zauner, \booktitle{Quantum Designs --
  Foundations of a Noncommutative Theory of Designs.} PhD thesis,
  University of Vienna
  (1999). \url{http://www.gerhardzauner.at/qdmye.html}.

\bibitem{Renes04} J.\ M.\ Renes, R.\ Blume-Kohout, A.\ J.\ Scott and
  C.\ M.\ Caves,
  ``\hrefdoi{10.1063/1.1737053}{Symmetric
    informationally complete quantum measurements},''
  \booktitle{Journal of Mathematical Physics} {\bf 45,} 6 (2004),
  2171, \arxiv{quant-ph/0310075}.

\bibitem{Scott10} A.\ J.\ Scott and M.\ Grassl,
  ``\hrefdoi{10.1063/1.3374022}{SIC-POVMs:\ A new
  computer study},'' \booktitle{Journal of Mathematical Physics} {\bf
  51,} 4 (2010), 042203, \arxiv[quant-ph]{0910.5784}.

\bibitem{Voldemort} C.\ A.\ Fuchs, ``QBism: The Perimeter of Quantum
  Bay\-esianism,'' \arxiv[quant-ph]{1003.5209} (2010).

\bibitem{RMP} C.\ A.\ Fuchs and R.\ Schack,
  ``\hrefdoi{10.1103/RevModPhys.85.1693}{Quantum-Bayesian
  Coherence},'' \booktitle{Reviews of Modern Physics} {\bf 85,} 4
  (2013), 1693--1715, \arxiv[quant-ph]{1301.3274}.

\bibitem{Fuchs2014} C.\ A.\ Fuchs and B.\ C.\ Stacey,
  ``Some negative remarks on operational approaches to quantum
  theory.'' In \booktitle{Quantum Theory: Informational Foundations
    and Foils} (Springer, 2016). \arxiv[quant-ph]{1401.7254}.

\bibitem{ConicalDesigns} M.\ A.\ Graydon and D.\ M.\ Appleby,
  ``\hrefdoi{10.1088/1751-8113/49/8/085301}{Quantum
  conical designs},'' \booktitle{Journal of Physics A} \textbf{49,} 8
  (2016), 085301, \arxiv[quant-ph]{1507.05323}.

\bibitem{RCF-SIC} D.\ M.\ Appleby, S.\ Flammia, G.\ McConnell and
  J.\ Yard, ``Generating ray class fields of real quadratic fields via
  complex equiangular lines,'' \arxiv[math.NT]{1604.06098} (2016).

\bibitem{Zhu2016} H.\ Zhu, ``Quasiprobability representations of
  quantum mechanics with minimal negativity,''
  \arxiv[quant-ph]{1604.06974} (2016).

\bibitem{zhu-thesis} H.\ Zhu, \booktitle{Quantum State Estimation and
  Symmetric Informationally Complete POMs.}  PhD thesis, National
  University of Singapore (2012). \\
  \url{http://scholarbank.nus.edu.sg/bitstream/handle/10635/35247/ZhuHJthesis.pdf}.

\bibitem{Szymusiak2015} A.\ Szymusiak and W.\ S\l{}omczy\'nski,
  ``Informational power of the Hoggar SIC-POVM,'' \\
  \arxiv[quant-ph]{1512.01735} (2015).

\bibitem{stacey-qutrit} B.\ C.\ Stacey, ``SIC-POVMs and Compatibility
  among Quantum States,'' \\ \arxiv[quant-ph]{1404.3774} (2014).

\bibitem{zhu2015} H.\ Zhu,
  ``\hrefdoi{10.1016/j.aop.2015.08.005}{Super-symmetric
    informationally complete measurements},'' \booktitle{Annals of
    Physics} \textbf{362} (2015), 311--26,
  \arxiv[quant-ph]{1412.1099}.

\bibitem{baez-review} J.\ C.\ Baez, ``\booktitle{On Quaternions and
  Octonions: Their Geometry, Arithmetic, and Symmetry} by John
  H.\ Conway and Derek A.\ Smith,'' \booktitle{Bulletin of the
    American Mathematical Society}
  \textbf{42} (2005),
  229--43. \url{http://math.ucr.edu/home/baez/octonions/conway_smith/}.

\bibitem{wilson2009} R.\ A.\ Wilson, \booktitle{The Finite Simple
  Groups} (Springer, 2009).

\bibitem{integral-octonions} J.\ C.\ Baez, G.\ Egan and T.\ Silverman,
  ``Integral Octonions,''
  \url{http://math.ucr.edu/home/baez/octonions/integers/} (2016).

\bibitem{ncatlab-ade} U.\ Schreiber, ``ADE Classification,''
  \url{https://ncatlab.org/nlab/show/ADE+classification} (2015).

\bibitem{viazovska2016} M.\ Viazovska, ``The sphere packing problem in
  dimension 8,'' \arxiv[math.NT]{1603.04246} (2016).

\end{thebibliography}
\end{document}